# Multiple magnetic states, valley electronics, and topological phase transitions in two-dimensional Janus *XYZ*H (*X* = Sc, Y, La, *Y* = Cl, Br, I, and *Z* = S, Se, Te): From monolayers to bilayers

Xinyu Tian[a], Zixuan Zhang[a], Lixiu Guan[a], Xiaobiao Liu[b], Xiaoyu Zhao[c,*], Linyang Li[a,**]

[a]*School of Science, Hebei University of Technology, Tianjin 300401, China*

[b]*School of Science, Henan Agricultural University, Zhengzhou 450002, China*

[c]*College of Materials and Environmental Engineering, Hangzhou Dianzi University, Hangzhou 310018, China*

*Corresponding author.

**Corresponding author.

E-mail address: zhaoxy@hdu.edu.cn (X. Zhao), linyang.li@hebut.edu.cn (L. Li)

**Abstract**

Exploring the coupling between layer, magnetism, valley, and topology in two-dimensional (2D) materials is an important approach to deepen our understanding of materials properties. We propose 27 stable ferromagnetic semiconductor monolayers of Janus *XYZ*H (*X* = Sc, Y, La, *Y* = Cl, Br, I, and *Z* = S, Se, Te). All these monolayers exhibit spontaneous valley polarization, forming ferrovalley (FV) monolayers, showing anomalous valley Hall (AVH) effect. By applying external strain, the topological phase transitions including quantum anomalous Hall (QAH) effect can be introduced. In the ScBrSH bilayer system, AA and AB stacking configurations were constructed through interlayer sliding and rotational operation. The bilayer system exhibits interlayer antiferromagnetic (AFM) ordering with spontaneous valley polarization differing from the FV observed in monolayers. The sliding ferroelectricity observed in the AA stacking indicates that the system exhibits significant multiferroic characteristics. Further analysis shows that interlayer sliding can introduce a layer polarization anomalous valley Hall (LPAVH) effect, which can be precisely controlled by tuning the direction of the ferroelectric polarization. Upon applying external strain, the quantum layer spin Hall (QLSH) effect observed during the topological phase transition in the bilayer can be regarded as the superposition of two QAH monolayers. Furthermore, applying a twisting operation to the bilayer induces unexpected altermagnetism. Our study systematically reveals the rich physical properties of 2D *XYZ*H materials, providing important theoretical foundations and guidance for the design and development of next-generation quantum devices.

In recent years, the valley degree of freedom has been regarded as an additional dimension beyond charge and spin, referring to the energy extrema at high-symmetry points in the valence or conduction band.[1] Experimentally, valley polarization has been induced through Zeeman splitting, such as, monolayers of $MoSe_2$,[2,3] and $WSe_2$[4,5] can be controlled by an external magnetic field. The spontaneous valley polarization observed in two-dimensional (2D) ferromagnetic (FM) material exhibits FM valley (FV) effect, showing great potential for applications in information encoding, transmission, and storage.[6-8] The existence of FV has been experimentally confirmed in $VS_2$ monolayer.[9] So far, we have found that most of the theoretically predicted FV monolayers focus on Fe-group compounds, such as $FeX_2$ ($X$ = Cl, Br, I),[10-13] $FeClBr$,[14] and $FeClF$,[15] $Ru(OH)_2$,[16] $RuCl_2$,[17] $RuBr_2$,[18,19] $RuClBr$,[20] $OsBr_2$,[21] and $OsClBr$.[22] The spontaneous valley polarization in FV monolayers mainly arises from strong spin-orbit coupling (SOC) and breaking of inversion symmetry. Besides the FV effect, applying external strain or tuning the electron correlation strength can introduce topological phase transitions, where the valley and quantum anomalous Hall (QAH) states should be coupled, offering new design possibilities for quantum devices.[23] A natural question arises: Can we find other transition-metal-group for the new family of FV monolayers? Based on the previous works, the rare-earth metals should also be a good choice.[24] Here, discovering 2D rare-earth metal compounds with FV effect and regulating their topological phase transitions through external strain or electron correlation effects is of great importance for advancing novel spintronic and topological quantum devices.

The layer stacking structure of 2D van der Waals (vdW) materials often leads to novel physical properties. By performing interlayer sliding and rotation operations on $CrI_3$ bilayers, the novel coexistence of FM and antiferromagnetic (AFM) orders has been confirmed.[25] These interlayer operations break the bilayer's inversion symmetry, inducing spontaneous valley polarization. The valley effect exhibited by this interlayer AFM order differs from the FV effect observed in monolayers. Similarly, applying external strain to the AFM bilayer FeCl$_2$[26] should induce a topological phase transition, and it was found that the monolayer QAH effect manifests as the quantum layer spin Hall (QLSH) effect.

In addition, stacking FM monolayers can couple magnetic materials with ferroelectric (FE) materials. For example, bilayers of Cr$_2$NO$_2$,[27] VS$_2$,[28] FeCl$_2$,[29] and VSi$_2$P$_4$[30] exhibit multiferroicity. Even more strikingly, through the interlayer operation, the spontaneous layer polarization anomalous valley Hall (LPAVH) effect has been realized in the AFM bilayer MnBi$_2$Te$_4$.[31] LPAVH effect combines the layer degree of freedom with opposite Berry curvature in real space, which is of significant importance for the development of novel spintronic devices.[26,32] Besides sliding manipulation, twisting has emerged as a new degree of freedom for tuning the properties of vdW bilayer. For example, studies have demonstrated that twisting can induce a unique altermagnetism in AFM bilayers.[33] Altermagnetism represents a novel magnetic state distinct from conventional FM and AFM orders.[34,35] It is characterized by collinear compensated magnetic ordering in real space and broken time-reversal symmetry in reciprocal space, accompanied by reversible chiral spin splitting.[36] A key feature of this altermagnetism is the momentum-dependent alternating spin splitting in the band structure, which occurs even in the absence of SOC.[35,37] This phenomenon originates from symmetry breaking and is closely associated with the crystal Hall effect.[38] Hence, the different properties can be found in magnetic states, valley electronics, and topological phase transitions from monolayer to bilayer, especially for the bilayer system composed of Janus monolayers.

Starting from the bulk hexagonal layered lattice of LaBr$_2$, which has a very low exfoliation energy, we can obtain monolayer and bilayer structures via mechanical exfoliation.[39,40] In this study, we constructed a Janus structure by substituting one layer of halogen atoms. Our focus is on 27 monolayers in the *XYZ*H family (*X* = Sc, Y, La, *Y* = Cl, Br, I, and *Z* = S, Se, Te). Each monolayer in this family exhibits FM ground state and spontaneous valley polarization due to the intrinsic breaking of inversion symmetry. Applying external strain can induce topological phase transitions and QAH states in these monolayers. We further investigated valley phenomena in bilayer stacking configurations, revealing the LPAVH/QLSH effect and altermagnetic order. This stacking configuration opens new avenues for investigating the rich physical properties of magnetism, multiferroicity, valley phenomenon, and topological phase transition.

We perform first-principles calculations in the framework of density functional theory (DFT) using the Vienna *ab-initio* Simulation Package (VASP) code.[41,42] Interactions between ions and valence electrons are treated using the projected augmented wave (PAW) method,[43,44] and electron exchange correlation functions are treated with Perdew, Burke and Ernzerhof (PBE) functionals within the generalized gradient approximation (GGA).[45,46] Atomic positions and planar lattice parameters were fully optimized using a conjugate gradient (CG) scheme until the maximum force on each atom was less than 0.01eV/Å. In the bilayer system, the vdW interactions between the layers are described using the DFT-D3 method.[47] In the structure optimization, the energy cut-off point of the plane wave basis set was set to 500 eV, and the energy accuracy was $10^{-5}$ eV, and a higher energy accuracy of $10^{-7}$ eV was used in other calculations. The Brillouin zone (BZ) was sampled using a 23 × 23 × 1 centered Monkhorst Pack grid. In all calculations, a vacuum separation of at least 20 Å was employed to minimize all interactions between two neighboring plates. In order to obtain a more accurate electronic band structure, we also adopted the Heyd-Scuseria-Ernzerhof (HSE06) hybrid function calculation method.[48,49] Phonon spectra are calculated using the supercell method in the PHONOPY code.[50,51] The GGA + $U$ method[52] was used to describe the correlated $X$-d and $X$-f electrons. The FE polarization is evaluated using the Berry phase approach, and the transition path and energy barrier of multiferroic materials are obtained by climbing the image nudge elastic band (c-NEB).[53] The Berry curvatures and edge states were calculated by the maximally localized Wannier functions as implemented in the wannier90 and wanniertools packages.[54,55]

The *XYZ*H ($X$ = Sc, Y, La, $Y$ = Cl, Br, I, and $Z$ = S, Se, Te) monolayer exhibits a hexagonal lattice and belongs to the space-group *P*3m1 (No.156). Fig. 1(a) illustrates the lattice structure composed of four layers of H-*Z*-*X*-*Y* atoms. The breaking of spatial inversion symmetry in the Janus system makes the conditions for the emergence of the valley effect. The detailed information about the lattice parameters of 27 monolayers in the *XYZ*H monolayer is summarized in Table 1. Notice that the lattice constants *a* of the monolayer become larger with increasing atomic number of the *X*/*Y*/*Z* atom. Furthermore, the dynamic stability of the 27 *XYZ*H monolayers can be confirmed by the phonon spectra, as shown in Fig. 1(d)-(f) and Fig. S1, where no imaginary frequencies can be observed.

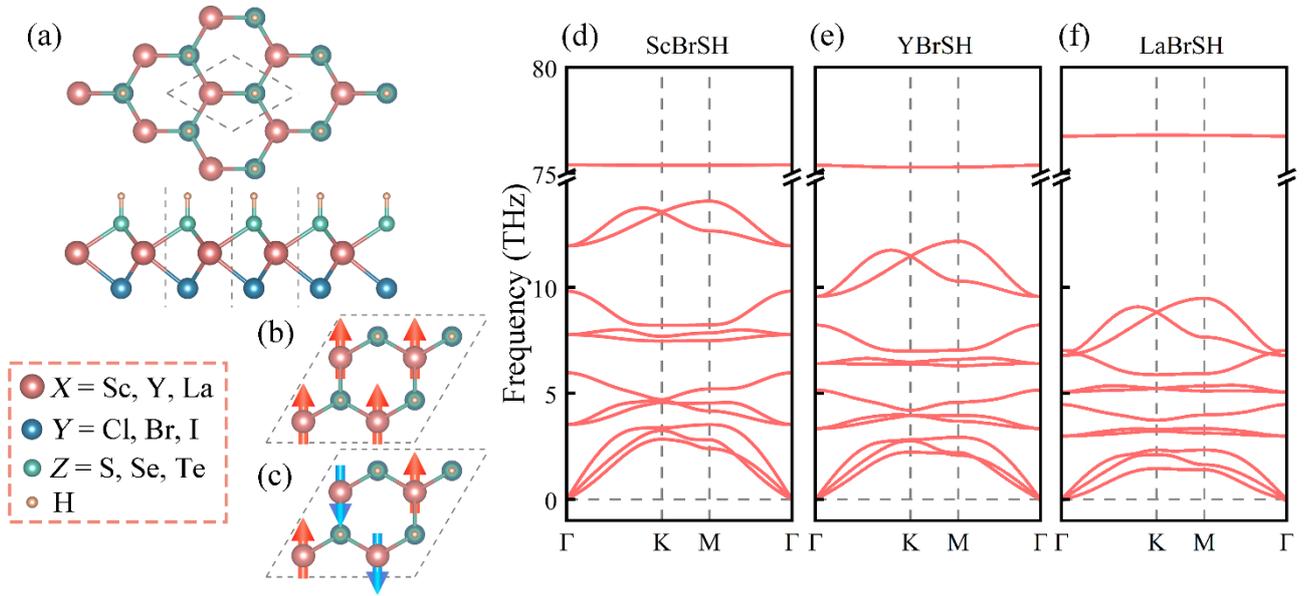

**Fig. 1** (a) Top and side views of the *XYZ*H monolayer. (b) FM and (c) AFM states of crystal structures, where the red (blue) arrow shows the direction of spin up (spin down). Phonon spectra of (d) ScBrSH, (e) YBrSH, and (f) LaBrSH along the high-symmetry paths.

**Table 1** Lattice constants $a$ (Å), bond length (Å), bond angle (°), MAE (μeV), $\Delta E$ (meV/unit cell), and $T_{BKT}$ (K).

| Monolayer | $a$ | Bond length | | | Bond angle | | MAE | $\Delta E$ | $T_{BKT}$ |
|---|---|---|---|---|---|---|---|---|---|
| | | $X$-$X$ | $X$-$Y$ | $X$-$Z$ | $X$-$Y$-$X$ | $X$-$Z$-$X$ | $E_z - E_x$ | $E_{AFM} - E_{FM}$ | |
| ScClSH | 3.67 | 3.67 | 2.65 | 2.59 | 87.8 | 90.4 | 10.49 | 63.58 | 246 |
| ScClSeH | 3.78 | 3.78 | 2.67 | 2.69 | 89.9 | 89.9 | 52.49 | 64.93 | 251 |
| ScClTeH | 3.94 | 3.94 | 2.72 | 2.87 | 92.9 | 86.6 | 267.72 | 56.99 | 221 |
| ScBrSH | 3.76 | 3.76 | 2.79 | 2.61 | 84.8 | 92.3 | 55.63 | 69.73 | 270 |
| ScBrSeH | 3.86 | 3.86 | 2.81 | 2.71 | 86.8 | 90.8 | 97.62 | 72.45 | 281 |
| ScBrTeH | 4.02 | 4.02 | 2.85 | 2.88 | 89.7 | 88.4 | 277.4 | 66.29 | 257 |
| ScISH | 3.90 | 3.90 | 2.97 | 2.64 | 82.0 | 95.2 | 169.46 | 76.04 | 295 |
| ScISeH | 4.00 | 4.00 | 2.99 | 2.74 | 83.8 | 93.6 | 203.06 | 82.33 | 319 |
| ScITeH | 4.15 | 4.15 | 3.03 | 2.91 | 86.4 | 90.9 | 237.88 | 57.75 | 224 |
| YClSH | 3.87 | 3.87 | 2.79 | 2.75 | 87.9 | 89.6 | 24.19 | 35.56 | 138 |
| YClSeH | 3.96 | 3.96 | 2.81 | 2.85 | 89.6 | 88.1 | 73.89 | 35.32 | 137 |
| YClTeH | 4.10 | 4.10 | 2.84 | 3.02 | 92.4 | 85.6 | 327.92 | 29.35 | 114 |
| YBrSH | 3.94 | 3.94 | 2.92 | 2.76 | 84.6 | 91.1 | 76.51 | 37.41 | 145 |
| YBrSeH | 4.03 | 4.03 | 2.94 | 2.86 | 86.3 | 89.6 | 149.67 | 37.21 | 144 |
| YBrTeH | 4.17 | 4.17 | 2.97 | 3.02 | 88.9 | 87.1 | 357.62 | 31.20 | 121 |
| YISH | 4.06 | 4.06 | 3.11 | 2.79 | 81.5 | 93.5 | 289.72 | 39.58 | 153 |
| YISeH | 4.14 | 4.14 | 3.13 | 2.88 | 83.1 | 92.0 | 438.47 | 40.74 | 158 |
| YITeH | 4.28 | 4.28 | 3.16 | 3.04 | 85.4 | 89.3 | 329.52 | 36.68 | 142 |
| LaClSH | 4.15 | 4.15 | 2.94 | 2.94 | 90.0 | 90.0 | 49.75 | 35.37 | 137 |
| LaClSeH | 4.23 | 4.23 | 2.95 | 3.03 | 91.5 | 88.4 | 136.2 | 36.40 | 141 |
| LaClTeH | 4.34 | 4.34 | 2.97 | 3.20 | 93.6 | 85.4 | 373.4 | 35.13 | 136 |
| LaBrSH | 4.21 | 4.21 | 3.08 | 2.94 | 86.2 | 91.2 | 114.53 | 36.31 | 141 |
| LaBrSeH | 4.28 | 4.28 | 3.09 | 3.04 | 87.7 | 89.5 | 208.96 | 37.15 | 144 |
| LaBrTeH | 4.39 | 4.39 | 3.11 | 3.20 | 89.9 | 86.7 | 493.02 | 35.28 | 137 |
| LaISH | 4.30 | 4.30 | 3.26 | 2.96 | 82.4 | 93.2 | 308.14 | 37.23 | 144 |
| LaISeH | 4.38 | 4.38 | 3.28 | 3.05 | 83.8 | 91.6 | 454.17 | 38.23 | 148 |
| LaITeH | 4.49 | 4.49 | 3.30 | 3.21 | 85.8 | 88.7 | 744.05 | 36.34 | 141 |

After confirming the dynamic stability of the *XYZ*H monolayer, we investigated the magnetic properties arising from the inclusion of rare-earth metal $X$ ($X$ = Sc, Y, La) atoms within the monolayer. We constructed a 2×2×1 supercell containing four $X$ atoms to determine the magnetic ground state of the monolayer. The FM and AFM states are shown in Fig. 1(b) and (c), respectively. The energy difference is calculated by using $E_{AFM} - E_{FM}$, where $E_{AFM}/E_{FM}$ represent the energy of the monolayer unit cell in the AFM/FM states, indicating that all monolayers are in the FM ground state. The magnetic moment of the monolayer is 1 $\mu_B$ per unit cell, mainly from the rare-earth metal $X$ atom. The FM coupling in the monolayer can be explained by the competitive mechanism of the direct exchange ($X$-$X$) and superexchange ($X$-$Y$/$Z$-$X$) interactions.[56] Taking the ScBrSH monolayer as an example, the distance between two adjacent Sc atoms is $d_{Sc-Sc}$ = 3.76 Å, and $d_{Sc-Br}/d_{Sc-S}$ = 2.79/2.61 Å. The large distance between adjacent Sc atoms indicates that the direct coupling interaction between Sc atoms is weak. Based on Goodenough-Kanamori-Anderson (GKA) rules,[57-59] when the cation-anion-cation angle approaches 90°, the superexchange interaction tends to FM coupling, while the 180° tends to AFM coupling. In the ScBrSH monolayer, the angles of Sc-Br-Sc/Sc-S-Sc bond is 84.8°/92.3°, which is close to 90°, indicating the FM coupling of the monolayer. To describe the magnetic anisotropy energy (MAE) of the monolayer, we calculated it by using MAE = $E_z - E_x$.[60] The result is a positive value, indicating that the easy magnetization axis of the monolayer should be in-plane. According to the Mermin-Wagner theorem, the monolayer exhibits quasi-magnetic long-range order,[61] and magnetic materials with in-plane MAE belong to $XY$ magnets.[62-64] The Berezinskii-Kosterlitz-Thouless (BKT) magnetic transition temperature $T_{BKT}$ of a 2D triangular $XY$ magnet can be estimated as $T_{BKT}$ = 1.335 $J/k_B$,[62-65] where $J$ and $k_B$ are the nearest-neighbor exchange parameter and the Boltzmann constant, respectively. The $J$ can be calculated through the energy difference between the FM and AFM states by using $J = (E_{AFM} - E_{FM})/4$.[64] The $T_{BKT}$ values are summarized in Table 1.

After confirming the magnetic properties of the monolayer, we further investigated its electronic band structure. We used the HSE06 and GGA+$U$ methods to calculate the electronic band structures without

SOC for the ScBrSH, YBrSH, and LaBrSH monolayers, as shown in Fig. S2. Based on the results of HSE06, the Hubbard $U$ parameters for the $d/d/f$ atomic orbitals of Sc/Y/La atoms should be 3/2/6 eV. Based on the above parameters, we calculated the electronic band structures of the monolayers without and with SOC along the $z$-direction, and the results are shown in Figs. S3 and S4, respectively. The calculations indicate that all these monolayers exhibit an indirect bandgap. Due to the breaking of central inversion symmetry in the monolayer 2H-Janus structure, the band structure with SOC exhibits energy non-degeneracy between the K and K' valleys, with the valley splitting ($E_K - E_{K'}$) marked in Fig. S4. Taking the ScBrSH monolayer as an example, we further analyzed the changes in its band structure under the effect of SOC, as shown in Fig. 2(d). With SOC, the valence band maximum (VBM) and conduction band minimum (CBM) of the ScBrSH monolayer exhibit 100% spin polarization with the spin polarization directions being opposite. These results exhibit bipolar magnetic semiconducting properties. The valley splitting of this monolayer is 46.3 meV, which is higher than the reported value of LaI$_2$ (28 meV) and PrI$_2$ (35 meV).[66] Due to the monolayer structure simultaneously breaking both spatial inversion and time-reversal symmetry, non-zero Berry curvature arises at the K and K' valleys. To further characterize the electronic transport properties of this monolayer, we calculated the Berry curvature using the following formula:[67]

$$\Omega_z(\bm{k}) = -\sum_n \sum_{m \neq n} 2 f_n(\bm{k}) \frac{\text{Im} <\psi_{nk}|\hat{v}_x|\psi_{mk}><\psi_{mk}|\hat{v}_y|\psi_{nk}>}{(E_{nk} - E_{mk})^2},$$

where $f_n(\bm{k})$ is the Fermi–Dirac distribution function, $\hat{v}_x$ and $\hat{v}_y$ are the velocity operators, and $E_{nk}$ and $E_{mk}$ are the eigenvalues of the Bloch wave functions $\psi_{nk}$ and $\psi_{mk}$, respectively. As shown in Fig. 2(g), the Berry curvature along the high-symmetry paths and within the 2D Brillouin zone were calculated. It can be observed that the Berry curvatures at the K and K' valleys exhibit opposite signs and different absolute values. Furthermore, we investigated the changes in the band structure and Berry curvature of the system after reversing the magnetic moment of the Sc atom and rotating the crystal structure by 180°. The band structure for the Sc magnetic moment along the negative $z$-direction (M↓) is shown in Fig. 2(e). Compared to the band structure with M↑ [Fig. 2(d)], the spin and valley directions are reversed, where the energy of the K' valley becomes higher than that of the K valley. Accordingly, the values of its Berry curvature [Fig. 2(h)] are exchanged, while the signs remain unchanged, indicating that the valley degree of freedom can be controlled by an external magnetic field. If the monolayer crystal structure is rotated by 180° while maintaining M↓, the resulting band structure [Fig. 2(f)] shows a reversal in the valley polarization sign compared to the unrotated case, where both the magnitude and sign of the Berry curvature are changed. Therefore, by reversing the direction of the magnetic moment or rotating the crystal structure by 180°, it is possible to tune the valley polarization and Berry curvature, providing new avenues for the control of spintronic and valleytronic devices.

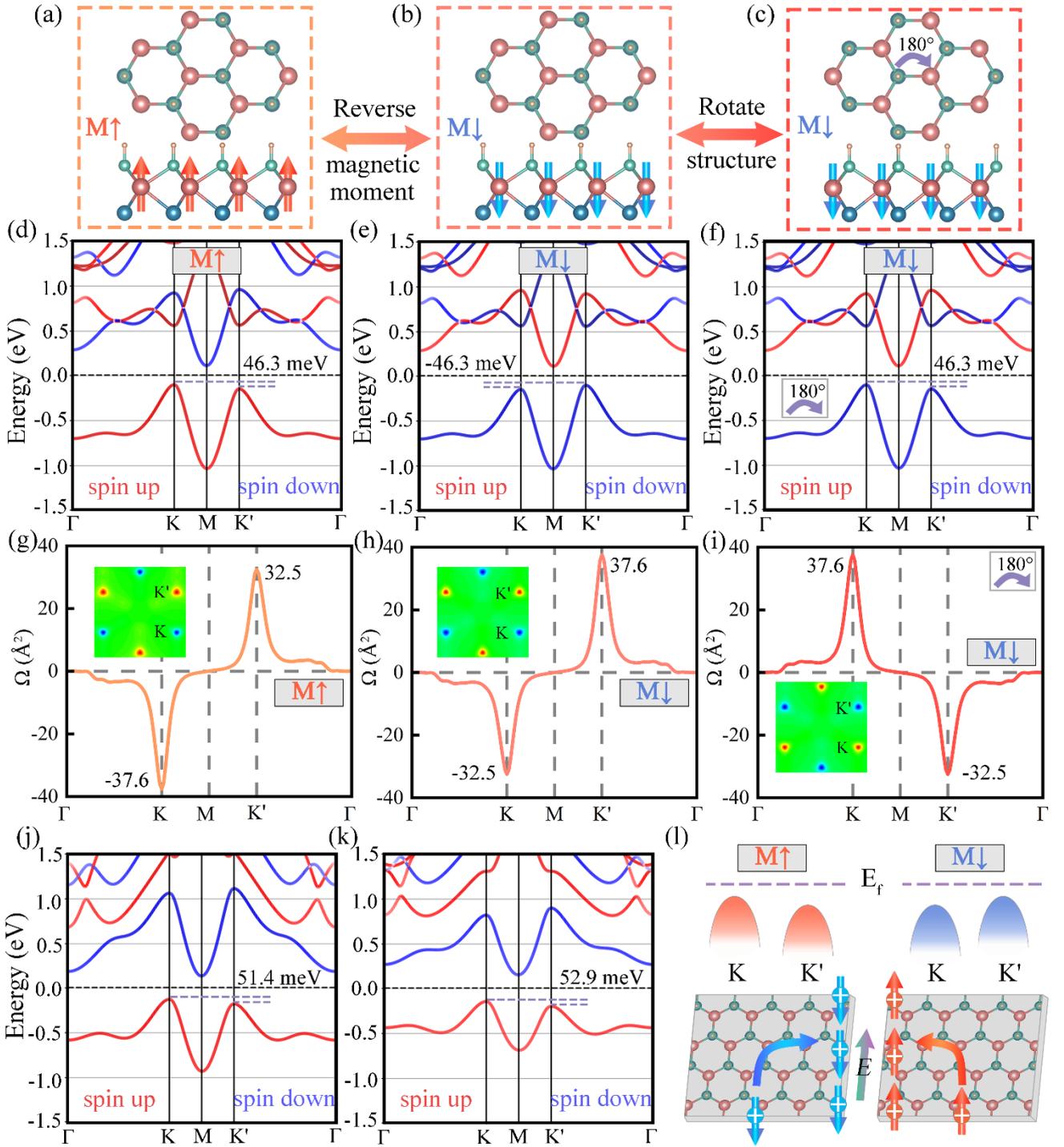

**Fig. 2** (a)-(c) Structures, (d)-(f) spin resolved band structures, and (g)-(i) Berry curvatures of the ScBrSH monolayer. (j) Spin resolved band structures of YBrSH monolayer and (k) LaBrSH monolayer. (l) Schematic diagram of AVH effect under external electric field *E*, where red and blue arrows represent spin up and spin down holes, respectively.

In addition, we also calculated the band structures of YBrSH and LaBrSH monolayers considering

SOC effect [Fig. 2(j) and (k)], and the results are similar to those of the ScBrSH monolayer. Under the influence of an in-plane electric field, the Bloch electrons acquire an anomalous velocity: $v_\perp = -\frac{e}{\hbar} E \times \Omega_z(k)$ .[68] By introducing hole doping, the position of the Fermi level can be adjusted, enabling the selective excitation of charge carriers in different valleys to accumulate on one side of the sample, thereby inducing an anomalous valley Hall (AVH) effect in the monolayer, as shown in Fig. 2(l). At the same time, by changing the magnetization direction of the rare-earth metal atom *X* to reverse the valley polarization, the Berry curvature signs between the energy valleys can be exchanged with maintaining the same values. It allows the AVH effect response to be controlled by detecting opposite-sign transverse voltages on either side of the sample, effectively enabling the adjustment of valley splitting through the manipulation of the magnetization direction. This approach can lead to new functional applications in spintronic devices.

The intrinsic AVH effect in the monolayer provides a valuable opportunity for further exploration of topological property control through external means. Additionally, the unique mechanical flexibility of 2D materials makes it relatively easier to tune material properties through strain in experimental settings. It is well known that strain is an effective means of tuning the electronic structure and topological properties of materials. To gain a deeper understanding of the strain-driven topological transition mechanism, we investigated the orbital-resolved band structures under force field [Fig. 3(a)]. Taking the ScBrSH monolayer as an example, strain is defined as $\varepsilon = (a - a_0)/a_0$, where *a* and $a_0$ represent the lattice constants with and without strain, respectively. It can be observed that external strain effectively changes the band structure and topological properties of the monolayer, while it consistently maintains a robust FM ground state under strain [Fig. S5]. The variation of the monolayer's bandgap with external strain is shown in Fig. 3(b). When the strain is less than 4%, both the valence and conduction bands at the K and K' points shift towards the Fermi level, resulting in a gradual decrease in the bandgap. Additionally, the K and K' valleys in the valence band exhibit inequivalent characteristics, leading to a FV state in the monolayer. As the external strain further increases to 5.25%, the bandgap at the K point closes, resulting in a gapless Dirac cone with linear dispersion, while the bandgap at the K' valley remains open. In this case, the ScBrSH monolayer exhibits half valley metal (HVM) characteristics, where electrons show high mobility. When the external strain slightly exceeds 5.25%, the previously closed bandgap at the K point reopens. Clearly, the $d_{z^2}$ orbital at the K point shifts from the CBM to the VBM, while the degenerate $d_{xy}$ and $d_{x^2-y^2}$ orbitals shift from the VBM to the CBM. The band structure at the K point remains unchanged, resulting in a single-valley band inversion phenomenon. When the external strain increases to 5.48%, the bandgap at the K' valley closes, while the bandgap at the K valley remains open. The K' valley undergoes a single-valley band inversion again. Subsequently, as the external strain further increases, the valence band will be entirely contributed by the $d_{z^2}$ orbital, while the conduction band will be completely contributed by the degenerate $d_{xy}$ and $d_{x^2-y^2}$ orbitals. The monolayer transitions back to the FV state, with the K and K' valleys shifting to non-degenerate valleys in the conduction band.

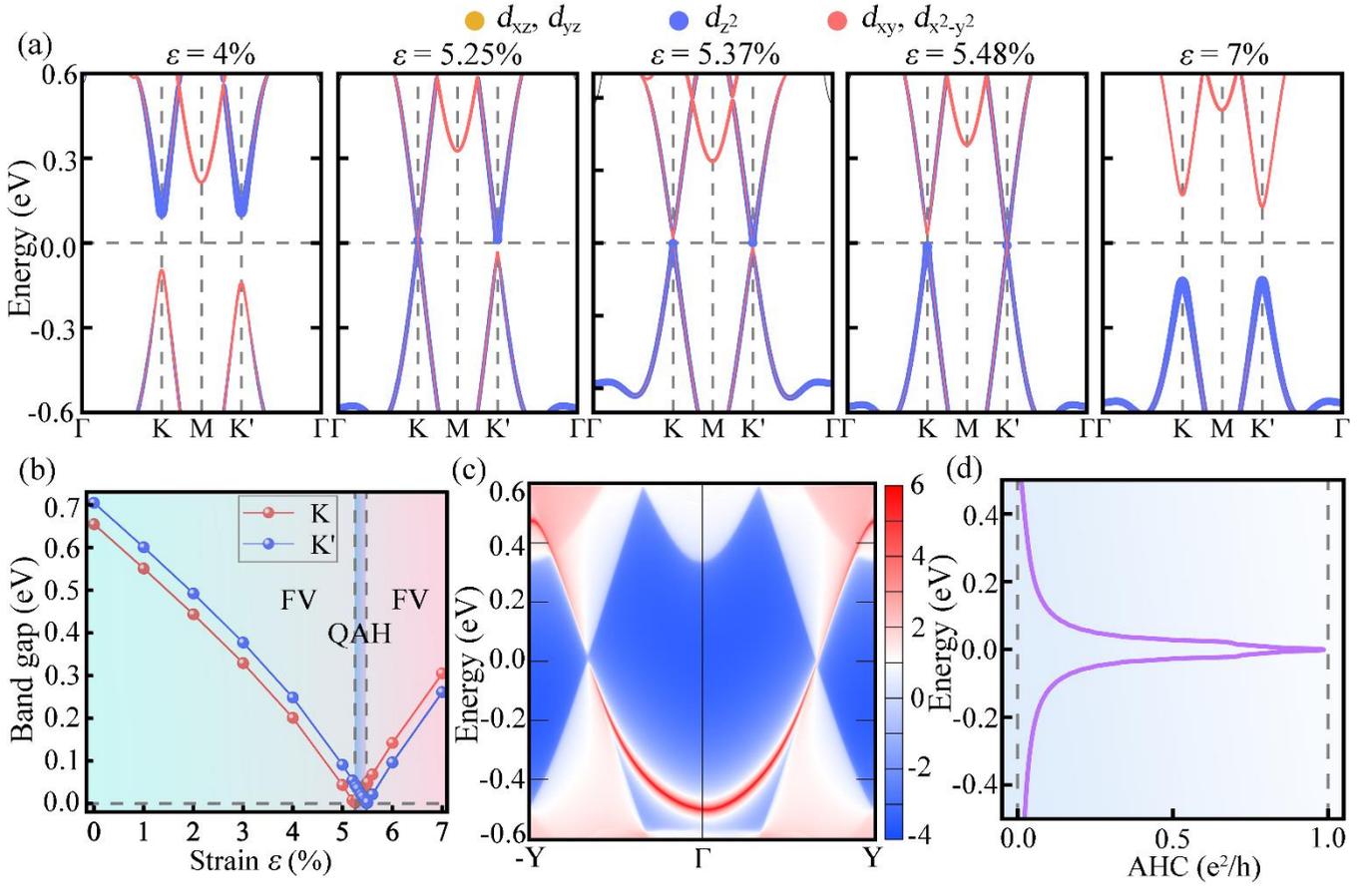

**Fig. 3** (a) Orbital-resolved band structure of the ScBrSH monolayer under strains, with the magnetization of the Sc atoms oriented in the +z direction. The band inversions occur sequentially at the K and K' points. (b) Variation of the bandgap at the K and K' valleys with strain. (c) Edge states and (d) AHC of the ScBrSH monolayer under 5.37% strain.

By applying external strain to the monolayer, the bandgaps at the K and K' valleys close and reopen at strain levels of 5.25% and 5.48%, respectively, accompanied by the single-valley band inversion phenomenon. It indicates that a strain-driven topological phase transition occurs. Within the strain between 5.25% and 5.48%, the monolayer exhibits a QAH state, and the anomalous Hall conductivity (AHC) and QAH edge states shown in Fig. 3(c) and (d). Additionally, the 2D AHC is calculated using the following formula:[69] $\sigma_{xy} = -\frac{e^2}{h}\int_{BZ}\frac{d\boldsymbol{k}}{(2\pi)^2}\Omega_z(\boldsymbol{k})$. Fig. 3(d) shows that the AHC in the bandgap is $e^2/h$, exhibiting a quantized value of the Chern number. The edge states create a conductive channel connecting the valence area and conduction area. This confirms that the system is a topologically nontrivial insulator with Chern number $C = 1$, corresponding to the QAH state. To gain physical insights into the strain-driven topological phase transitions, we also calculated the Berry curvature before and after the two instances of band inversion [Fig. S6]. It can be observed that when the K and K' valleys undergo band inversion successively, and the corresponding signs of the Berry curvature also reverse.

Layer stacking in 2D magnetic materials plays a crucial role in expanding their physical properties, introducing novel effects that are unattainable in monolayers. Here, we take the ScBrSH bilayer as an example and consider two stacking structures, namely AA and AB, as shown in Fig. 4(a). In the AA stacking, the AA-0 bilayer is obtained by simply placing one layer on top of another with the Sc atoms in the bottom layer corresponding to the Sc atoms in the top layer. The AA stacking can be obtained from the AA-0 stacking through interlayer sliding operations of $[-1/3, -1/3, 0]$ or $[1/3, 1/3, 0]$. It is worth noting that the loss of mirror symmetry in the AA-1 stacking induces the emergence of out-of-plane FE polarization. Under interlayer translation operations, the AA-2 stacking can be obtained from the AA-1 stacking, and it also exhibits out-of-plane FE polarization. The AB stacking presents an inversion symmetry, resulting in the absence of spontaneous FE polarization. By rotating the top layer of the AA-0 stacking by 180°, we achieved the AB-0 stacking and two additional configurations AB-1 and AB-2 can be derived through interlayer sliding. The interlayer magnetism of the bilayer structure can be switched through interlayer sliding or rotation. To understand the interlayer magnetic behavior of these bilayers, we evaluated the energies of six stacking configurations for both FM and AFM ordering. As shown in Fig. 4(a), the interlayer AFM is energetically preferred over FM ordering for all the six cases. To ensure dynamical stability, we calculated the phonon spectra for all six stacking configurations, confirming stability across these configurations.

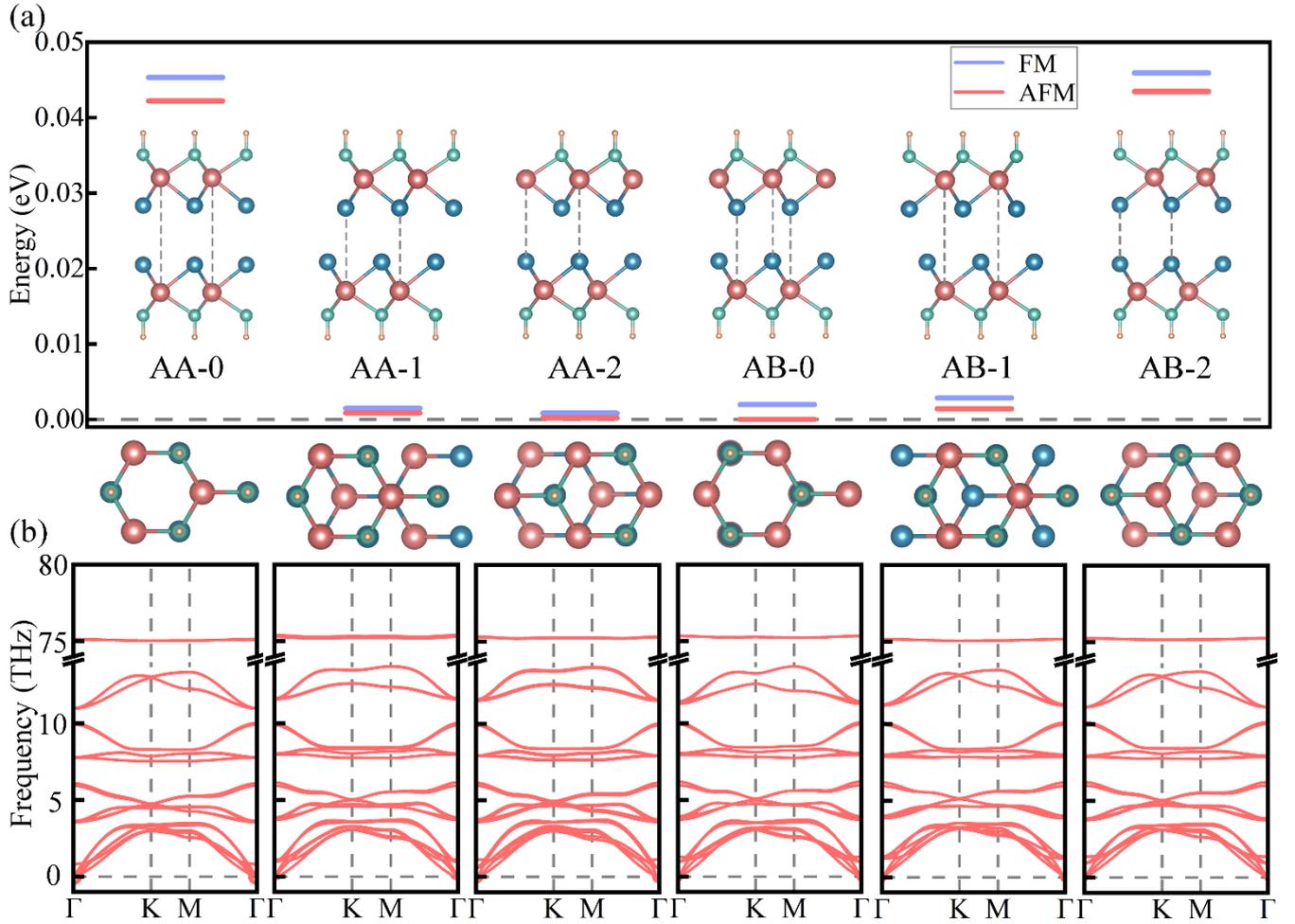

**Fig. 4** (a) Side and top views of AA-type and AB-type stacked ScBrSH bilayers with a comparison of the energies of interlayer FM and AFM orders. (b) Phonon spectra corresponding to the AA-type and AB-type stacked ScBrSH bilayers.

According to the energies of the six stacking structures marked in Fig. 4(a), the AA-0 and AB-2 stacking types show relatively high energies, while the other four stacking modes (AA-1, AA-2, AB-0, and AB-1) have lower energies. We further investigated the sliding energy barriers of two stacking structures, as shown in Fig. S7(a). The AA-1 and AA-2 exhibit relatively low energies, and an energy barrier of 11.9 meV is required to transition from the AA-1 stacking to the AA-2 stacking via an intermediate state. For the AB stacking, AB-2 exhibits the highest sliding energy of 42 meV, as shown in Fig. S7(b). Due to the broken structural symmetry along the z-direction, the AA-1 and AA-2 stacking structures are expected to exhibit out-of-plane FE polarization. To confirm this property, we calculated the spontaneous polarization values of AA-1 and AA-2 stacking using the Berry phase method, obtaining −42.7 pC/m and 42.7 pC/m for AA-1 and AA-2 stacking, respectively. Notably, the AA-1 and AA-2 configurations have the same energy and exhibit opposite out-of-plane FE polarization, as indicated by the arrows in Fig. 5(a). Therefore, the AA-1 and AA-2 stackings of ScBrSH bilayers are FE materials with an FE switching barrier of 11.9 meV, similar to that of the $YI_2$ bilayer (13.63 meV).[70] Notably, the AA-1 and AA-2 stacking configurations exhibit both AFM and FE properties, showing the characteristic of typical multiferroic behavior.

Based on the AFM properties of the bilayer structure, we investigated the electronic band structures of AA-1 and AA-2 stackings under SOC with the M↓↑ magnetic configuration [Fig. 5(b) and (c)]. The valence and conduction bands show spin layer locked polarization, where the AA-1 (AA-2) configurations shows an indirect bandgap. Additionally, the degeneracy at the K and K' valleys is broken, resulting in valley polarizations of −10 meV (10 meV) in the valence band for AA-1 (AA-2). This valley polarization of two stacked states can be reversed. The reversal of valley polarization between the K point (spin up channel in the bottom layer) and the K' point (spin down channel in the top layer) indicates that valley polarization can be tuned via interlayer sliding. The calculated Berry curvature in Fig. 5(d) and (e) further confirms the reversal of valley polarization. For AA-1 stacking, the Berry curvatures at the K and K' valleys are −60.1 $Å^2$ and 60.2 $Å^2$, respectively, while for AA-2 stacking, they are −60.2 $Å^2$ and 60.1 $Å^2$, respectively. Obviously, by sliding the stacking from AA-1 to AA-2, the signs of the Berry curvature at the K and K' valleys remain unchanged, while the absolute values are exchanged. We can conclude that the layer-locked Berry curvature from the locking relationship between the valleys and layers. In the AA-1 stacking of ScBrSH bilayers, applying an in-plane electric field and hole doping can tune the Fermi level between the K and K' valleys. Under these conditions, holes at the K' valley can gain transverse velocity and accumulate on one side of the sample, inducing LPAVH effect.[71] In contrast, in the AA-2 configuration, the holes at the K valley acquire opposite transverse velocity, accumulating at the bottom layer's left boundary. It is worth noting that LPAVH effect has been observed in both AA-1 and AA-2 stacked bilayers. This indicates that the LPAVH effect observed in the ScBrSH bilayer can be controlled and reversed through FE switching, which is significant for the external regulation of spin device functionalities.

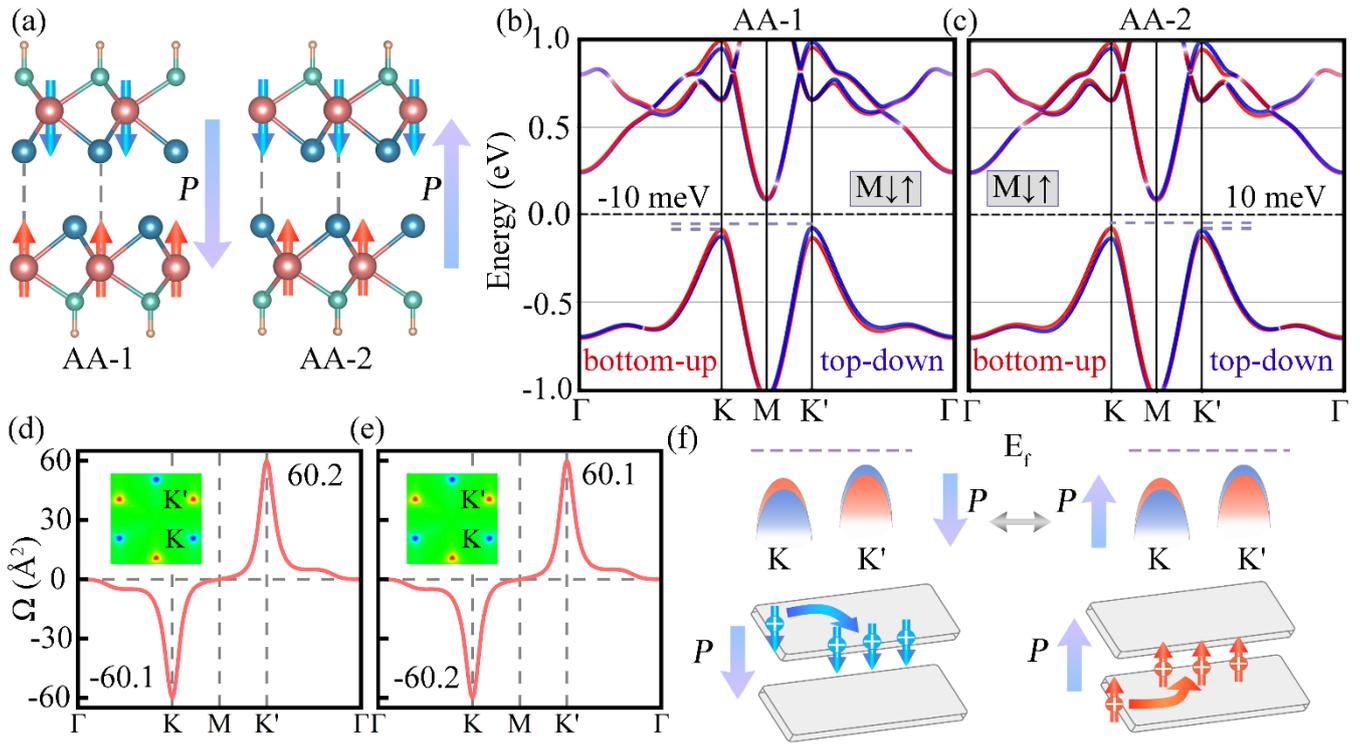

**Fig. 5** (a) Top and side views of AA-1 and AA-2 stacked ScBrSH bilayer structures where the arrows indicate the direction of spontaneous polarization. Electronic band structure with SOC for (b) AA-1 and (c) AA-2 stacking. Berry curvature of the ScBrSH bilayer of (d) AA-1 stacking and (e) AA-2 stacking. (f) Schematic diagram of LPAVH effect.

Unlike the AA stacking, the AB stacking with inversion symmetry exhibits complete overlap between the spin up and spin down bands [Fig. S8(b)-(d)]. This overlap has a significant impact on valley polarization, particularly displaying distinct characteristics in the AB stacking. The AB stacking exhibits a valley polarization of approximately 47 meV, similar to the value of the monolayer. Interestingly, this valley polarization cannot be reversed through interlayer sliding. When the magnetic moment direction changes from M↓↑ to M↑↓, as shown in Fig. S8(c) and (e), both the spin orientation and the sign of the valley polarization are reversed. Furthermore, in the AB stacking, the Berry curvature of the top and bottom layers is equal in magnitude but opposite in sign, and therefore the overall Berry curvature of the bilayer can be neglected in space. Hence, LPAVH effect can be observed in the AA-1 and AA-2 stacked ScBrSH bilayers, while it is clearly absent in the AB stacking.

As previously discussed, we examined how strain induces a topological phase transition in the ScBrSH monolayer. Based on this, we explored the effect of strain on the electronic and topological properties of the ScBrSH bilayer. Specifically, we applied strain to the stable AA-1/AB-1 stacking configurations of the AA/AB structure to investigate the evolution of their topological properties. Under a tensile strain of up to 7%, the magnetic ground state remains stable with an interlayer AFM order. Fig. 6(a) shows the evolution of the band structure for the AB-1 configuration under strain. The variation of the bandgap at the K and K' points with strain shows a similar trend to that in the monolayer, as shown in Fig. 6(c). When the external strain is below 6.10%, the K and K' valleys in the valence band exhibit inequivalent characteristics, resulting in the ScBrSH bilayer displaying a valley state. As the strain increases, the bandgaps at the K and K' valleys close and reopen at strains of 6.10% and 6.35%, respectively, where a band inversion occurs. The bilayer transition from the valley state to the HVM state. The band inversion indicates that a topological phase transition occurs in the response of each layer. As strain continues to increase, the bandgap opens, and the bilayer returns to the conventional valley state. The corresponding edge states are shown in Fig. 6(b). When $\varepsilon = 6.20\%$, the edge states exhibit two conductive channels in different directions. Considering the spin layer locking feature in the ScBrSH bilayer, these two channels can be understood as arising from the fully decoupled QAH insulator contributions of the top and bottom layers. Furthermore, due to the opposite magnetization directions between the two layers, the channels exhibit opposite chirality. The coupling of the QAH effect and layer degrees of freedom gives rise to the QLSH effect, exhibiting transport phenomena similar to those of the quantum spin Hall (QSH) effect. This phenomenon has been rarely reported in previous studies, highlighting its unique significance in the field of topological physics. Using the Kubo-Greenwood formula, we calculated the spin Hall conductivity (SHC) with the following equation:[72]

$$\sigma_{xy}^{s_z}(\omega) = \hbar \int_{BZ} \frac{d^3k}{(2\pi)^3} \sum_n \sum_{m \neq n} 2 f_n(\boldsymbol{k}) \frac{\mathrm{Im} <\psi_{nk}|\hat{j}_x^{s_z}|\psi_{mk}><\psi_{mk}|-e\hat{v}_y|\psi_{nk}>}{(E_{nk}-E_{mk})^2 - (\hbar\omega + i\eta)^2},$$

where $\hat{j}_x^{s_z}$ represents the projection of the spin current operator in the z-direction. Under a direct current environment, both the angular frequency ω and damping factor η are reduced to zero. We calculated the quantized SHC value at the strain of 6.20% to be −2, in units of e/4π, indicating that there is no coupling between the layer-polarized edge states and the bulk states. The value of SHC can be defined as $\sum_{l=1}^{2}(-1)^l C_l$,[73] where $C_l$ is the layer Chern number. We found that the AFM state QLSH insulator is obtained by stacking two QAH insulators with equally but oppositely oriented magnetic moments. Similarly, this theory is also applicable to the AA-1 stacking configuration of the ScBrSH bilayer. Fig. S9 shows that the phenomenon observed in the AA-1 stacking is similar to that in the AB-1 stacking. The SHC value calculated for the AA-1 stacking is also close to −2, as shown in Fig. S9(c). The AFM state QLSH insulator is characterized by a spin chiral layer locked edge state, opening up vast opportunities for investigating a novel quantum material and spin transport mechanism.

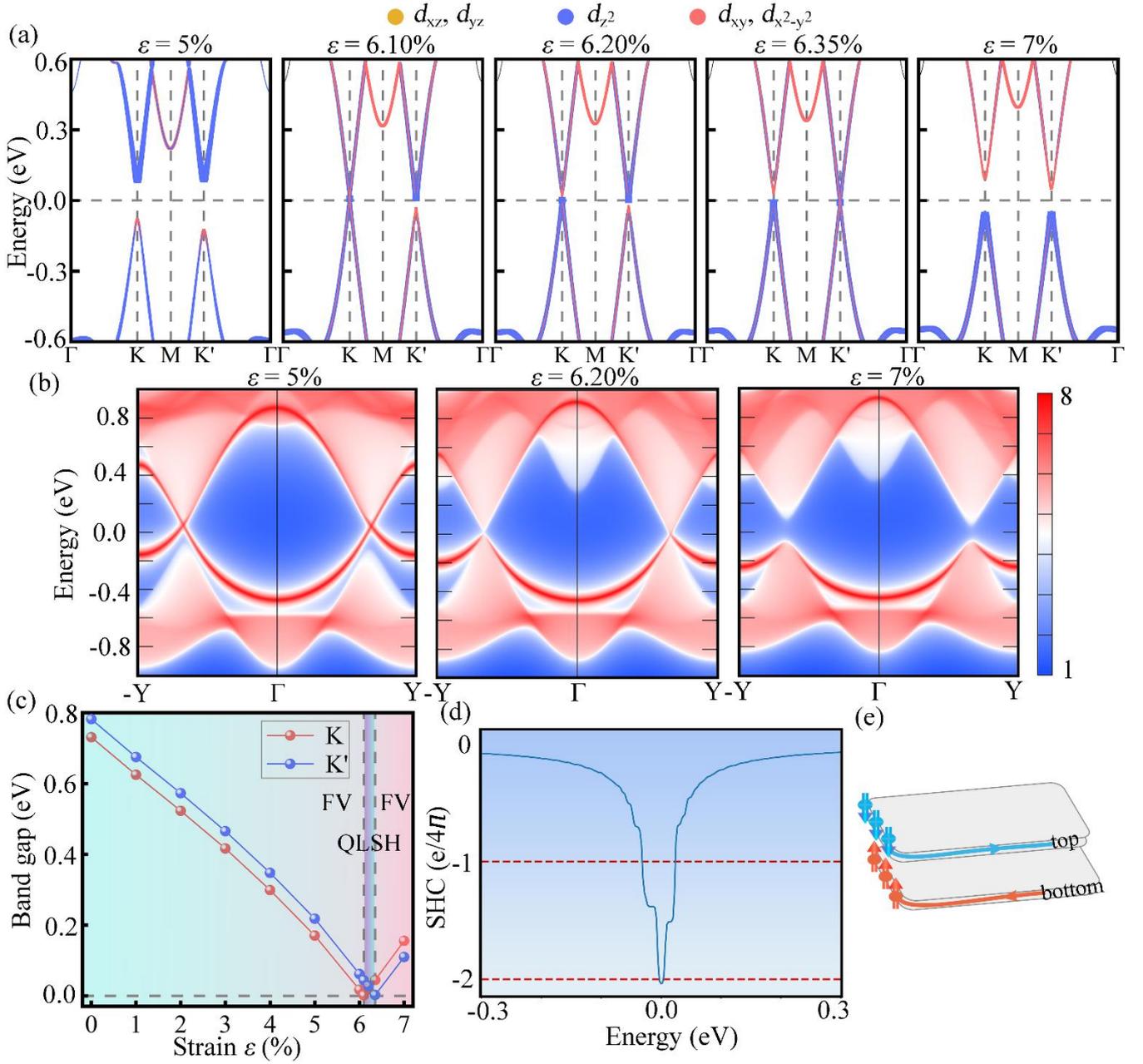

**Fig. 6** (a) Orbital projected band structures for the AB-1 stacking ScBrSH bilayer considering SOC under different strains. (b) Topological edge states under tensile strain $\varepsilon$ = 5%, 6.20%, and 7%. (c) Variation of the bandgap at the K and K' valleys with tensile strain. (d) SHC under $\varepsilon$ = 6.20%. (e) Schematic illustration for the spin chirality layer locked edge states of the AFM QLSH insulators, where red and blue colors represent the spin up and spin down stacked QAH insulators, respectively.

The above discussion demonstrates the effectiveness of interlayer sliding in tuning the properties of vdW stacked structures with a simple stacking model (AA/AB). Besides, the bilayer structure with a vdW interaction should lead to a complex stacking model, moiré structure. The realization of altermagnetism can be produced by the moiré stacking, which requires two key conditions: the crystal must exhibit compensated collinear magnetic ordering, and the opposite spins of the sublattices must be linked through rotational symmetry.[74] Applying twisting operations to bilayer vdW materials with interlayer AFM ordering has become a common method for constructing altermagnetism states.[75-77] Bilayers with a rectangular lattice possessing $C_2$ symmetry, such as VOBr,[36,37] Ca(CoN)$_2$,[37] and CrOCl,[78] show band splitting along high-symmetry paths. Materials with a square lattice possessing $C_4$ symmetry, such as CrN,[34] and Co$_2$S$_2$,[36] exhibit band splitting along the X-Y path, but spin degeneracy is maintained along the Γ-X and Y-Γ high-symmetry paths due to magnetic symmetry. Bilayers with a hexagonal lattice possessing $C_3$ symmetry, such as MnPSe$_3$,[33] NiCl$_2$,[34,78] PtBr$_3$,[35] and MnBi$_2$Te$_4$,[36] display band splitting along non-high-symmetry paths in the same environment, while maintaining degeneracy along the high-symmetry paths. Next, we investigate the mechanism of altermagnetism by applying twisting operations to ScBrSH bilayers. Among the six stacking configurations, the AB-0 stacking with the lowest total energy was chosen as the initial structure, and a twist was applied only to the top layer, forming the twisted ScBrSH bilayer structure, as shown in Fig. 7(a). In ScBrSH bilayer with AFM order, flipping the top layer introduces an in-plane rotational operation. The in-plane rotation couples the two layers with opposite spins, thereby giving rise to alternating magnetic states.[36] The electronic band structure without considering SOC is presented in Fig. 7(b). Along the non-high-symmetry path K-K$_1$, a pronounced spin splitting between spin up and spin down bands is observed, while along the high-symmetry paths Γ-K and K$_1$-Γ, the spin states remain degenerate. This altermagnetism arises from symmetry breaking induced by the twist. Specifically, the spin degeneracy along high-symmetry paths is protected by the spin layer group symmetry. Therefore, twisting can induce altermagnetism in 2D vdW bilayers, which holds significant implications for research and applications in spintronics in low-dimensional systems.

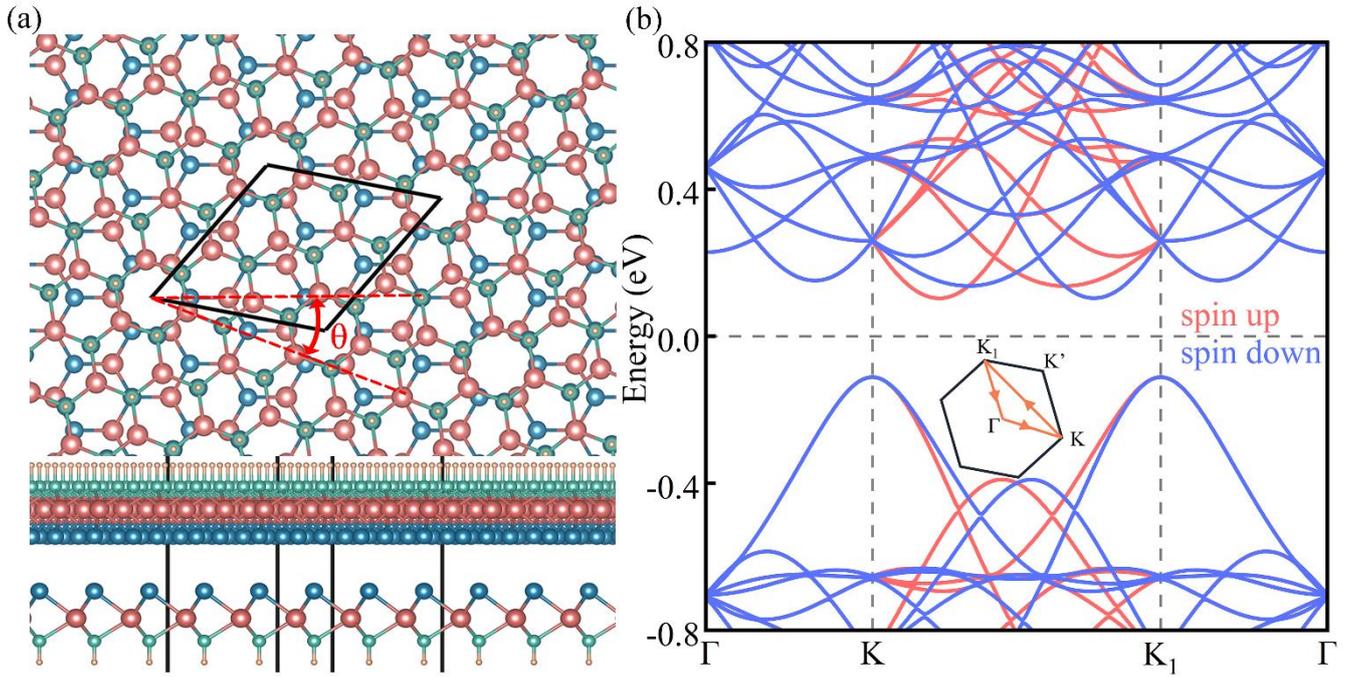

**Fig. 7** (a) Top and side views of the optimized twisted structure for the (2 1/1 2) bilayer ScBrSH with a rotation angle $\theta$ of 21.79° and a supercell containing 56 atoms. (b) Band structure of the twisted bilayer ScBrSH without considering SOC, where the illustration shows the first BZ and the positions of the high-symmetry points.

In summary, we have proposed a series of novel 2D FM semiconductors, Janus $XYZ$H ($X$ = Sc, Y, La, $Y$ = Cl, Br, I, and $Z$ = S, Se, Te) monolayers and bilayers, exhibiting rich physical properties including spontaneous valley polarization, multiferroic behavior, AVH/LPAVH effect, QAH/QLSH effect, and altermagnetism. These findings open up new avenues for tuning material properties through external fields. In the ScBrSH monolayer, the intrinsic FV and strain-driven topological states can be found, where the AVH and QAH effects can be realized. From the monolayer to the bilayer, two kinds of stacking models were applied. For the simple stacking model, we controlled bilayer stacking structure through interlayer sliding and rotational operation. In the AA stacking configuration, the coupling between interlayer AFM properties and FE polarization leads to the locking of Berry curvature, leading to LPAVH effect. Different from the monolayer, the application of external strain can introduce QLSH effect in the AFM bilayer stacking, where the band structure is composed of two QAH states with opposite chirality. The edge states exhibit spin-helical-locking characteristics and demonstrate a quantized SHC. For the complex stacking model of moiré structure, introducing twisting operations can realize the exciting features of altermagnetism in the bilayer. These theoretical findings highlight the tremendous potential of 2D Janus *XYZ*H materials in next-generation electronics and spintronics technologies.

**Acknowledgments**

This work was supported by the National Natural Science Foundation of China (Grant No. 12004097). X.Z. acknowledges the financial support from the National Natural Science Foundation of China (Grant No. 12374345, U24A20103), and Key R&D Project of Zhejiang Province (No. 2024C03258).